\documentclass[RNAAS]{aastex631}

\usepackage{amsmath}
\usepackage{hyperref}
\RequirePackage[fixed]{fontawesome5}
\newcommand{\github}[1]{%
   \href{#1}{\faGithub}%
}
\shorttitle{SpectralUnmix}
\shortauthors{de Souza}

\begin{document}

\title{SpectralUnmix: A Torch-Based Regularized Non-negative Matrix Factorization}

\author[0000-0001-7207-4584]{Rafael S. de Souza}
\email{r.s.desouza@herts.ac.uk}
\affiliation{Centre for Astrophysics Research, University of Hertfordshire, College Lane, Hatfield, AL10~9AB, UK}
\affiliation{Instituto de Física, Universidade Federal do Rio Grande do Sul, Porto Alegre, RS 90040-060, Brazil}
\affiliation{Department of Physics \& Astronomy, University of North Carolina at Chapel Hill, NC 27599-3255, USA}

\author[0000-0003-1846-4826]{Paula Coelho} 
\affiliation{Instituto de Astronomia, Geofísica e Ciências Atmosféricas, USP, Rua do Matão 1226, 05508-090, São Paulo, Brazil}

\author[0000-0003-3372-3638]{Niranjana P} 
\affiliation{Instituto de Física, Universidade Federal do Rio Grande do Sul, Porto Alegre, RS 90040-060, Brazil}

\author[0000-0000-0000-0002]{Ana L. Chies-Santos} 
\affiliation{Instituto de Física, Universidade Federal do Rio Grande do Sul, Porto Alegre, RS 90040-060, Brazil}

\author[0000-0002-1321-1320]{Rog\'erio Riffel}
\affiliation{Instituto de Física, Universidade Federal do Rio Grande do Sul, Porto Alegre, RS 90040-060, Brazil}

\begin{abstract}
We present \texttt{SpectralUnmix}, an \texttt{R} package for regularized non-negative matrix factorization (NMF), implemented in \texttt{torch} with optional GPU acceleration. The package estimates low-rank non-negative representations through proximal-gradient updates and allows smoothness regularization along the spectral axis. As a compact demonstration, we apply the method to a subset of stellar spectra and compare the recovered NMF components with principal-component directions and representative stellar spectra. The package is released under the MIT license at \href{https://rafaelsdesouza.github.io/SpectralUnmix/}{this repository \faGithub}.
\end{abstract}

\section{Introduction}

Many spectral datasets can be represented as matrices in which rows correspond to samples and columns correspond to wavelength channels. A common objective in exploratory analysis is to describe such data through a small number of latent components and their corresponding weights. 
Principal component analysis \citep[PCA;][]{Jolliffe2016} has long been used in astronomy for this purpose \citep[e.g.][]{Ronen1999,Steiner2009,ishida2013,Joseph2014,desouza2014,desouza2022}, providing an orthogonal basis that captures the dominant variance in spectral datasets.
Non-negative matrix factorization (NMF) offers an alternative decomposition in which both the components and their weights are constrained to be non-negative \citep{Lee1999}. Because astronomical spectra represent non-negative flux measurements, imposing non-negativity often yields components that are more directly interpretable in physical terms \citep[e.g.][]{Blanton2007,Hurley2013,Melchior2018}.  Such representations arise naturally in applications  ranging from stellar libraries to spatially resolved spectroscopy.
Our \texttt{R} package provides a general-purpose implementation of regularized NMF with \texttt{torch} as the computational backend. It is designed for exploratory decomposition problems where smooth, non-negative spectral components are desirable.

\section{Method}

Let $\mathbf{X}$ be a non-negative data matrix $\mathbf{X}\in\mathbb{R}_{+}^{N\times P}$,
where \(N\) denotes the number of samples and \(P\) the number of spectral channels. We seek a low-rank non-negative representation
$\mathbf{X}\approx \mathbf{A}\mathbf{S}$,
with $\mathbf{A}\in\mathbb{R}_{+}^{N\times K}$, and
$\mathbf{S}\in\mathbb{R}_{+}^{K\times P}$.
The rows of \(\mathbf{S}\) define \(K\) latent spectra, while the columns of \(\mathbf{A}\) contain the corresponding non-negative weights.
We estimate \(\mathbf{A}\) and \(\mathbf{S}\) by solving
\begin{equation}
\min_{\mathbf{A},\mathbf{S}\ge0}
\frac{1}{2}\left\|\mathbf{X}-\mathbf{A}\mathbf{S}\right\|_F^2
+
\lambda \left\|\mathbf{S}\mathbf{D}^{\top}\right\|_F^2,
\label{eq:objective}
\end{equation}
where $\mathbf{D}\in\mathbb{R}^{(P-1)\times P}$ is the first-order finite-difference operator along the spectral axis, 
\begin{equation}
\mathbf{D} =
\begin{bmatrix}
-1 & 1 & 0 & \dots & 0 \\
0 & -1 & 1 & \dots & 0 \\
\vdots & & & \ddots & \\
0 & 0 & 0 & -1 & 1
\end{bmatrix}, 
\end{equation}
so that the regularization penalizes large bin-to-bin variations and encourages smooth spectral components. In this case
\begin{equation}
\left\|\mathbf{S}\mathbf{D}^{\top}\right\|_F^2
=
\sum_{k=1}^{K}\sum_{j=1}^{P-1}
\left(S_{k,j+1}-S_{k,j}\right)^2 .
\end{equation}
The first term in \autoref{eq:objective} measures reconstruction fidelity, while the second penalizes channel-to-channel roughness in the recovered spectra. The standard NMF is recovered when \(\lambda=0\). The optimization is implemented in \texttt{torch} via alternating proximal-gradient updates \citep{Xu2013}.

\section{Analysis}

\begin{figure}[h]
\centering
\plotone{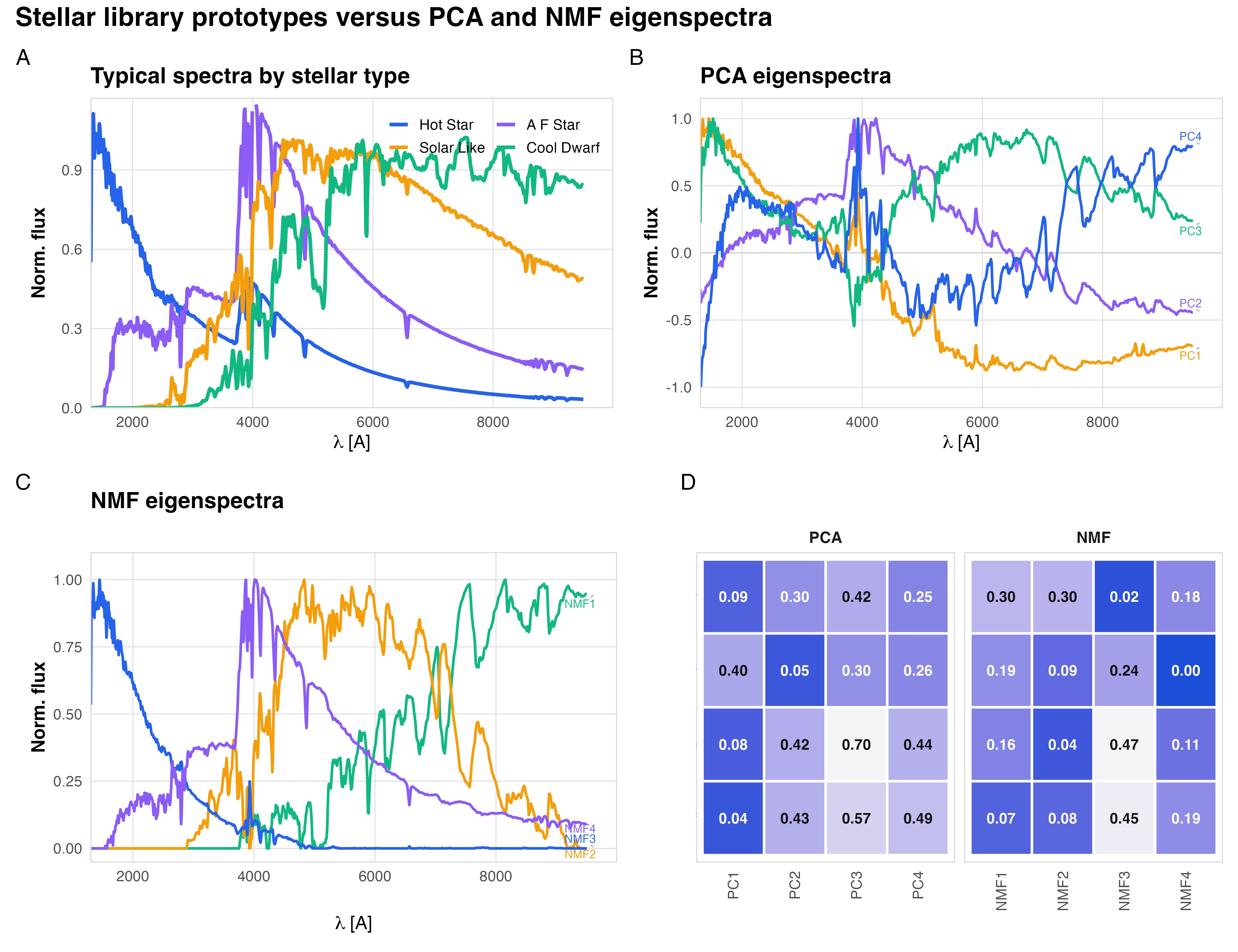}
\caption{Panel A shows four representative normalized spectra out a sample of 80. Panels B and C compare PCA and NMF eigenspectra derived from the same data; colors indicate the stellar class assigned to each component through a one-to-one matching based on a simple \(\chi^2\)-like distance between normalized spectral shapes. Panel D summarizes the class--component distances for PCA and NMF.}
\label{fig:mosaic}
\end{figure}

Here we apply \texttt{SpectralUnmix} to a subset of 80 representative stellar spectra from the spectral library of \citet{Coelho2014}, spanning four broad classes: hot stars, A/F stars, solar-like stars, and cool dwarfs, with 20 spectra per class. We fit a four-component NMF model and compare the recovered components with PCA obtained from the same data. \autoref{fig:mosaic} summarizes this toy experiment. Panel~A shows the class prototype spectra; panels~B and~C display the PCA and NMF eigenspectra, respectively, with colors indicating their one-to-one assignment to the prototype classes. For each component--class pair we compute the $\chi^2$ distance between the corresponding eigenspectrum and the class prototype. Finally, panel~D summarizes the class--component distances. In this example, the recovered NMF components trace physically recognizable spectral shapes while remaining smooth and non-negative by construction.

\section{Conclusions}

\texttt{SpectralUnmix} provides an extensible implementation of regularized NMF for spectral data analysis using \texttt{torch} in \texttt{R}. The package supports GPU acceleration and optional smoothness constraints, making it suitable for exploratory decomposition of high-dimensional spectral datasets. Although demonstrated here on a simple stellar library example, the framework is general and can be readily applied to a wide range of astronomical problems, including stellar population studies, hyperspectral imaging, and integral-field spectroscopy.

\begin{acknowledgments}
RSS acknowledges support from the Conselho Nacional de Desenvolvimento Cient\'ifico e Tecnol\'ogico (CNPq, Brazil), process nos. 446508/2024-1 and 315026/2025-1.
\end{acknowledgments}

\bibliography{spectralunmix}{}
\bibliographystyle{aasjournal}

\end{document}